\documentclass[final,5p,twocolumn,authoryear]{elsarticle}

\usepackage{amsmath,amsfonts,amssymb}
\usepackage{algorithmic}
\usepackage{algorithm}
\usepackage{array}
\usepackage{subfig}
\usepackage{textcomp}
\usepackage{stfloats}
\usepackage{url}
\usepackage{verbatim}

\journal{ISPRS Journal of Photogrammetry and Remote Sensing}
\biboptions{round,semicolon}

\begin{document}

\begin{frontmatter}

\title{A Task-Driven and Quality-Assured Agent Framework for SAR Data Generation}

\author[buct]{Xuanting Wu}
\author[buct]{Fan Zhang\corref{cor1}}
\ead{zhangf@mail.buct.edu.cn}
\author[buct]{Fei Ma}
\author[casic207]{Ling Guan}
\author[buct]{Guochun Ma}
\author[buct]{Yongsheng Zhou}

\affiliation[buct]{organization={College of Information Science and Technology, Beijing University of Chemical Technology},
            city={Beijing},
            postcode={100029},
            country={China}}

\affiliation[casic207]{organization={Science and Technology on Electromagnetic Scattering Laboratory, Beijing Institute of Environmental Features},
            city={Beijing},
            postcode={100854},
            country={China}}

\cortext[cor1]{Corresponding author.}

\begin{abstract}
Synthetic aperture radar (SAR) data augmentation is important for improving the generalization of data-driven SAR interpretation models, yet practical augmentation workflows are often hindered by heterogeneous dataset formats, task-dependent metadata requirements, diverse generation methods, and weak validation of generated samples. This paper presents the \textbf{S}AR \textbf{A}ugmentation and \textbf{G}eneration \textbf{A}gent (SAGA), a schema-grounded and benefit-aware agent framework for task-oriented SAR data generation and augmentation. Given a natural-language request and heterogeneous SAR inputs, SAGA extracts observable dataset facts, validates executable dataset schemas, selects feasible augmentation strategies through validator-constrained planning, and compiles the selected strategy into an auditable augmentation workflow. Generated data are further assessed by quality, distribution, SAR-artifact, duplicate, leakage, and optional downstream-task evaluators to support evidence-qualified augmentation claims. By separating semantic proposal from deterministic validation and execution, SAGA improves the reliability and reproducibility of SAR augmentation decisions. Experiments on controlled agentic benchmarks and downstream SAR interpretation tasks show that SAGA improves schema grounding, skill planning, invalid-sample rejection, and downstream augmentation utility compared with rule-based, LLM-only, ReAct-style, and fixed-augmentation baselines.
\end{abstract}


\end{frontmatter}

\section{Introduction}

Synthetic aperture radar (SAR) is an important sensing modality for remote sensing, surveillance, and target interpretation because of its all-weather and day-and-night imaging capability. In recent years, deep learning has achieved remarkable progress in SAR automatic target recognition (ATR), object detection, and image interpretation. However, data-driven SAR models remain highly dependent on the availability, diversity, and reliability of training data. Unlike optical images, SAR images are governed by radar imaging geometry and electromagnetic scattering mechanisms. Target appearance may vary significantly with azimuth angle, depression angle, polarization, frequency band, resolution, sensor platform, and background clutter. As a result, SAR models trained with limited or biased data often suffer from poor generalization under unseen acquisition conditions.

\begin{figure}[!t]
    \centering
    \includegraphics[width=0.9\linewidth]{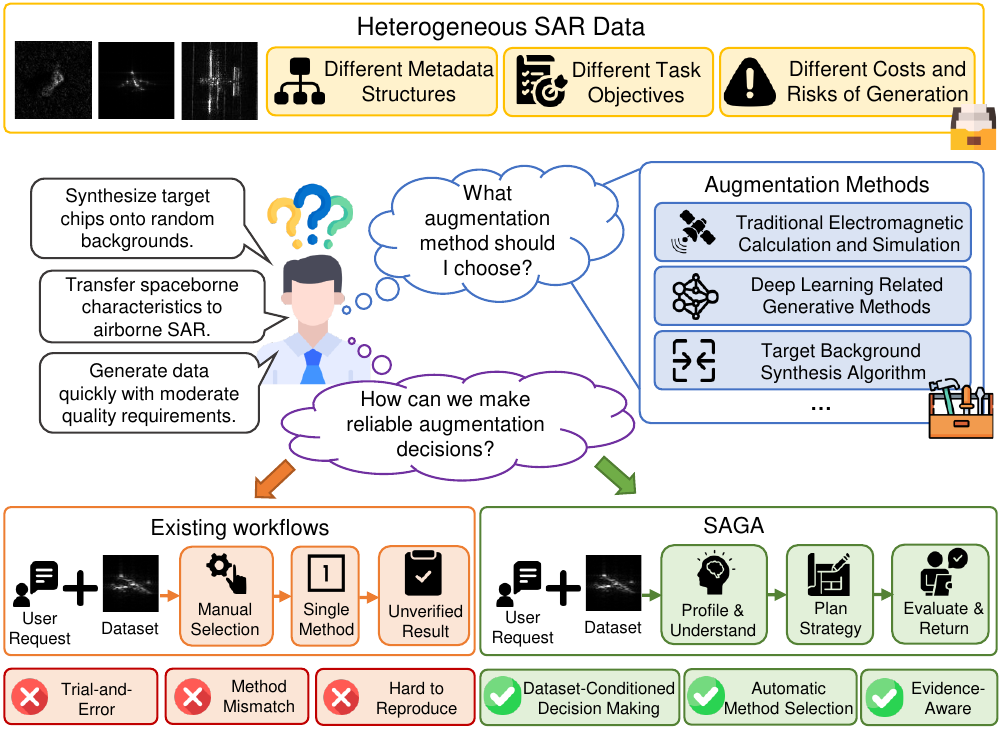}
    \caption{Motivation and problem landscape of SAGA. SAR data augmentation involves heterogeneous datasets, diverse user objectives, and augmentation methods with different assumptions, costs, and risks. Existing workflows typically rely on manual or fixed method selection, whereas SAGA formulates augmentation as a dataset-conditioned and evidence-aware decision process.}
    \label{fig:motivation}
\end{figure}

Data augmentation is a practical way to alleviate SAR data scarcity. Conventional transformations such as flipping, cropping, rotation, intensity perturbation, and noise injection are efficient, but they provide limited semantic diversity and may violate SAR-specific imaging characteristics if applied indiscriminately. Meanwhile, recent generative models, including generative adversarial networks (GANs), diffusion models, LoRA-adapted text-to-image models, and condition-guided generation models, have provided new possibilities for SAR image synthesis and domain adaptation~\citep{goodfellow2014gan,radford2016dcgan,ho2020ddpm,rombach2022latent,hu2022lora,zhang2023controlnet}. Physics-based SAR simulation and ray-tracing tools can further generate interpretable SAR signatures from 3D models~\citep{balz2015potentials,auer2016raysar}. Nevertheless, these augmentation tools have different assumptions, costs, benefits, and risks. Traditional augmentation is fast but limited, GAN-based generation may suffer from mode collapse, diffusion-based generation is expressive but computationally expensive, and physical simulation is interpretable but often affected by simulation-to-real domain gaps.

\begin{figure*}[t]
    \centering
    \includegraphics[width=0.95\textwidth]{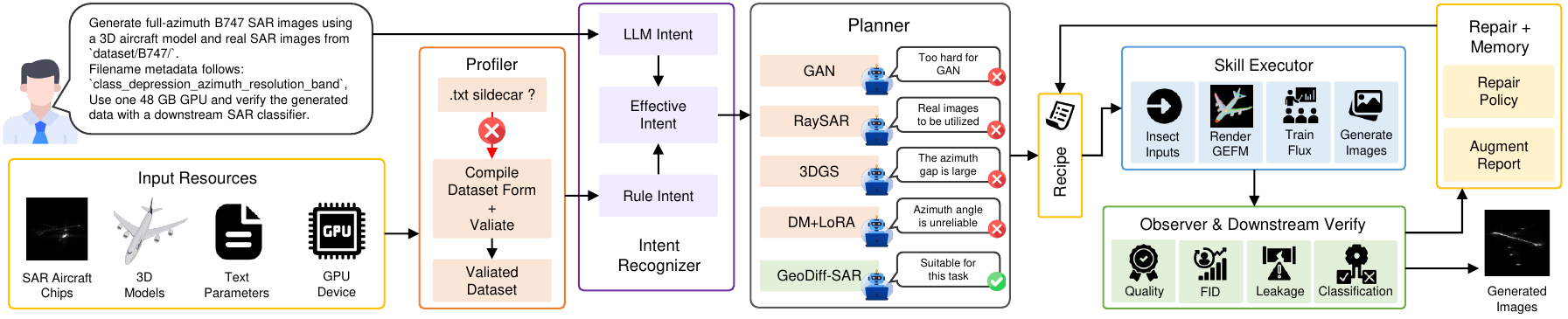}
    \caption{Concrete application scenario of SAGA for 3D-guided full-azimuth B747 SAR generation. Given a natural-language request, a 3D aircraft model, real SAR images, and filename-encoded metadata, SAGA performs schema-grounded profiling, intent recognition, guarded skill planning, recipe execution, observer verification, downstream classification evaluation, and bounded repair or memory update when necessary.}
    \label{fig:specific_example}
\end{figure*}

The central challenge is therefore not merely to generate more SAR images, but to make reliable augmentation decisions under heterogeneous dataset formats, natural-language user requirements, diverse augmentation tools, limited computational resources, and downstream task objectives. In realistic research workflows, SAR metadata such as category, polarization, azimuth, depression angle, resolution, and frequency band may be stored in directory names, file names, sidecar files, captions, or customized annotation formats. A fixed-format assumption is often too restrictive, while directly relying on a large language model (LLM) to infer dataset semantics and execute augmentation workflows may lead to unreliable or non-reproducible results. Dataset facts, metadata coverage, label consistency, filtering validity, and execution safety should instead be verified through deterministic procedures.

Figure~\ref{fig:motivation} summarizes the motivation and problem landscape. It contrasts manual or fixed augmentation workflows with the proposed SAGA paradigm, where augmentation decisions are conditioned on dataset structure, task requirements, method risks, and evaluation evidence.

To address these issues, we propose \textbf{SAGA}, a \textbf{S}AR \textbf{A}ugmentation and \textbf{G}eneration \textbf{A}gent for task-oriented dataset enhancement. SAGA is not a single SAR generator or a simple collection of augmentation scripts. Instead, it formulates SAR data augmentation as a schema-grounded and benefit-aware decision process. Given a user request and a SAR dataset, SAGA first performs deterministic dataset profiling to extract objective file-system, image, annotation, and metadata facts. It then combines user-provided format hints and optional LLM-assisted schema induction to compile a validated dataset schema, which converts heterogeneous external datasets into an internal executable representation. Based on the validated dataset profile, task intent, resource constraints, augmentation skill descriptors, profiling evidence, and policy memory, SAGA selects suitable augmentation strategies and compiles them into executable augmentation recipe DAGs.

Figure~\ref{fig:specific_example} illustrates a concrete SAGA execution trace for a 3D-guided full-azimuth B747 SAR generation task. In this example, the user provides a B747 3D model, real SAR chips, filename-encoded imaging parameters, a 48~GB GPU budget, and a downstream classification requirement. SAGA first grounds the filename schema into validated metadata fields, then recognizes the task as geometry-conditioned full-azimuth generation, selects a GeoDiff-SAR-centered recipe under compatibility and resource guardrails, executes the recipe, and finally attaches observer and downstream classification evidence to the exported augmented dataset.

\begin{figure*}[t]
    \centering
    \includegraphics[width=0.95\textwidth]{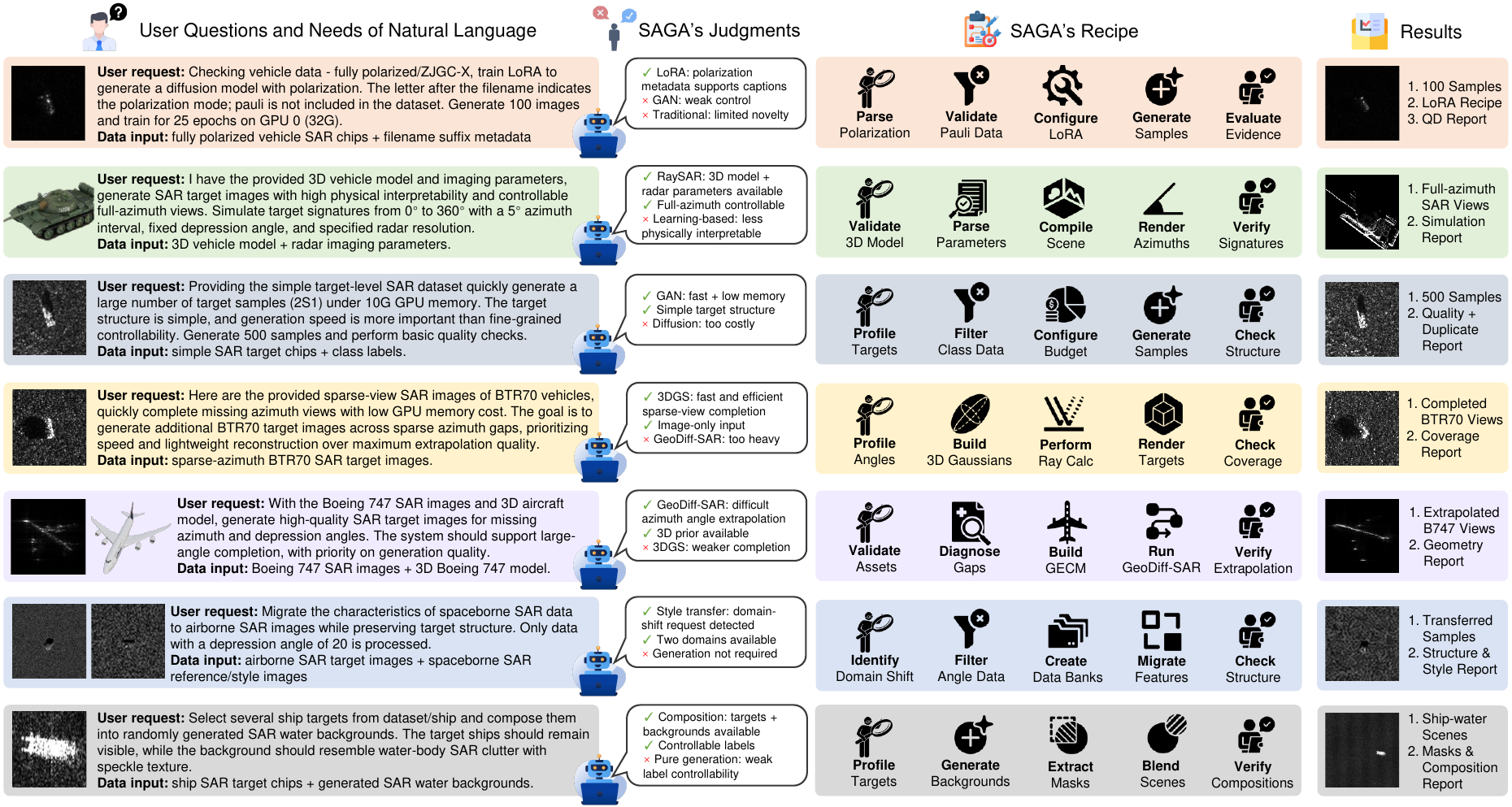}
    \caption{Examples of SAGA plans under different user requests and input data types. Given heterogeneous inputs such as polarimetric SAR chips, 3D object models, sparse-view target images, platform-specific references, and target--background sets, SAGA parses the request, evaluates candidate strategies, compiles case-specific recipes, and returns generated results with evidence reports.}
    \label{fig:case_gallery}
\end{figure*}

A central design principle of SAGA is to separate flexible planning from reliable execution. The LLM may assist in natural-language intent parsing, schema proposal, and high-level planning, but its outputs are restricted to proposals that must be verified before execution. Candidate plans are checked by deterministic guardrails, including format validators, skill compatibility rules, plan verification, and risk assessment. The final recipe is executed by a deterministic runtime, and generated data are evaluated by observer modules such as image quality checks, distribution probes, SAR artifact analysis, duplicate or near-duplicate detection, leakage checks, and optional downstream task evaluators.

Representative planning cases are shown in Figure~\ref{fig:case_gallery}. These examples illustrate that SAGA does not apply a fixed augmentation pipeline; instead, it selects different recipes for polarization-conditioned LoRA generation, physics-based simulation, lightweight GAN generation, sparse-view completion, geometry-guided extrapolation, style transfer, and target--background composition according to the input data and the user request.

The main contributions of this work are summarized as follows:
\begin{itemize}
    \item We propose a schema-grounded profiling and format-bridging mechanism for heterogeneous SAR datasets, which combines deterministic raw profiling, user format hints, LLM-assisted schema induction, and validator-gated dataset schema compilation.

    \item We design a benefit-aware hybrid planning framework for SAR data augmentation, which integrates task intent, dataset needs, resource constraints, augmentation capability descriptors, profiling evidence, and policy memory to generate executable recipe DAGs.

    \item We introduce an observer-driven evidence and bounded repair mechanism that distinguishes generation success from downstream task utility. The proposed framework uses quality, distribution, SAR artifact, duplicate, leakage, and optional downstream evaluators to prevent unsupported augmentation claims and accumulate reusable augmentation experience.
\end{itemize}

By shifting the focus from individual augmentation operations to dataset-conditioned augmentation decision making, SAGA establishes a reproducible and evidence-aware workflow for SAR data generation and augmentation.

\section{Related Work}

\subsection{SAR Augmentation and Recognition}

Data augmentation has long been used to improve the robustness of SAR ATR and interpretation models under limited training data. Conventional augmentation strategies, including cropping, translation, rotation, flipping, intensity perturbation, and speckle-like noise injection, are computationally efficient and can improve recognition stability in small-sample settings~\citep{furukawa2017saraugmentation,geng2023saratr}. Surveys of SAR ATR also show that modern recognition pipelines frequently rely on augmentation, transfer learning, and synthetic data to alleviate data scarcity~\citep{kechagias2020saratr_survey}. However, such transformations provide only limited semantic diversity and may alter SAR-specific structures such as scattering centers, shadows, layover patterns, and angle-dependent signatures. Automated augmentation policies developed for natural images, such as AutoAugment, RandAugment, Mixup, CutMix, and AugMix~\citep{cubuk2019autoaugment,cubuk2020randaugment,zhang2018mixup,yun2019cutmix,hendrycks2020augmix}, are therefore not directly sufficient for SAR augmentation. This motivates augmentation workflows that are conditioned on SAR metadata, task objectives, and modality-specific validity constraints.

\subsection{Generative and Physics-Based SAR Synthesis}

Generative models have expanded the space of SAR data synthesis beyond local image transformations. GAN variants~\citep{goodfellow2014gan,radford2016dcgan,gulrajani2017wgangp} and image-to-image translation models such as CycleGAN~\citep{zhu2017cyclegan} have been used for SAR target generation, small-sample ATR, and domain adaptation~\citep{cui2019sar_gan,kong2021sar_gan,oghim2024dhgan}. More recent SAR-specific generators introduce stronger controllability, such as azimuth-conditioned target image generation~\citep{wang2023azimuthgan}. Diffusion models have also been adapted to SAR image synthesis and augmentation, with studies exploring conditional and unconditional DDPM-based SAR generation~\citep{qosja2024sardiff} and broader surveys discussing the role of generative AI in SAR interpretation~\citep{huang2024genai_sar}. Despite their potential, generated SAR samples still require validation, because visual plausibility does not guarantee metadata consistency, physical interpretability, label correctness, or downstream utility.

Physics-based simulators provide a complementary route by generating SAR signatures from geometric and electromagnetic assumptions. Comparative studies have examined the capabilities and limitations of SAR simulators such as RaySAR, CohRaS, and SARViz~\citep{balz2015potentials}, and RaySAR provides an open-source 3D ray-tracing framework for generating SAR image layers from detailed object models~\citep{auer2016raysar}. Simulation-based data can support ATR and geometry-aware augmentation, but the resulting images often require refinement or adaptation to reduce simulation-to-real gaps~\citep{cha2018simulated_sar,feng2024singlescene_sar}. Overall, existing SAR generation and simulation methods provide useful capabilities, but each method has specific input requirements, costs, risks, and validity assumptions. SAGA builds on these capabilities by treating them as selectable augmentation skills rather than as a single fixed pipeline.

\subsection{Remote Sensing and Geospatial Agents}

LLM-based agents have recently been introduced into remote sensing and geospatial analysis. Remote Sensing ChatGPT connects ChatGPT with visual models for remote sensing task planning and execution~\citep{guo2024rschatgpt}. RS-Agent integrates remote sensing tools and retrieval-augmented knowledge for professional remote sensing questions~\citep{xu2024rsagent}. Change-Agent supports interactive change detection, captioning, object counting, and change interpretation~\citep{liu2024changeagent}. GeoGPT and GeoAgent extend LLM-based tool use to geospatial data collection, processing, mapping, and automatic geospatial analysis~\citep{zhang2024geogpt,chen2024geoagent}. ThinkGeo further evaluates tool-augmented agents on remote sensing tasks requiring structured tool use, argument binding, and multi-step reasoning~\citep{shabbir2025thinkgeo}. These works demonstrate the promise of LLM-driven tool coordination, but they primarily target image interpretation, question answering, change analysis, or geospatial processing. They do not specifically address SAR data augmentation, where the agent must ground heterogeneous dataset schemas, choose among multiple generation and simulation skills, execute reproducible recipes, and qualify downstream benefit claims.

\subsection{Tool-Augmented LLM Agents and Evaluation}

General-purpose tool-augmented agents provide foundations for the design of SAGA. ReAct interleaves reasoning and acting for interactive problem solving~\citep{yao2023react}, HuggingGPT uses an LLM as a controller to coordinate external models~\citep{shen2023hugginggpt}, ToolLLM studies large-scale API use~\citep{qin2024toolllm}, and AutoGen supports multi-agent conversation frameworks~\citep{wu2023autogen}. Recent benchmarks further highlight the importance of executable evaluation and rule adherence. AgentBench evaluates LLM-as-agent behavior across multiple environments~\citep{liu2024agentbench}, MLAgentBench evaluates agents on end-to-end machine-learning experimentation tasks~\citep{huang2023mlagentbench}, and $\tau$-bench tests tool agents under domain-specific policies and user interactions~\citep{yao2024taubench}. These studies suggest that tool-augmented agents should be evaluated not only by final answers but also by tool selection, argument correctness, execution validity, and policy compliance.

SAGA follows this direction but introduces constraints specific to SAR augmentation. It adopts LLM-assisted semantic proposal while relying on deterministic schema validation, skill guardrails, recipe-centric execution, observer evidence, bounded repair, and policy memory. In contrast to existing SAR augmentation methods that focus on individual generators and existing agents that focus on interpretation or general tool use, SAGA formulates SAR augmentation as a schema-grounded and evidence-aware decision process over heterogeneous datasets, diverse augmentation skills, execution constraints, and downstream utility requirements.

\section{Method}
\label{sec:method}

\begin{figure*}[htbp]
    \centering
    \includegraphics[width=0.95\textwidth]{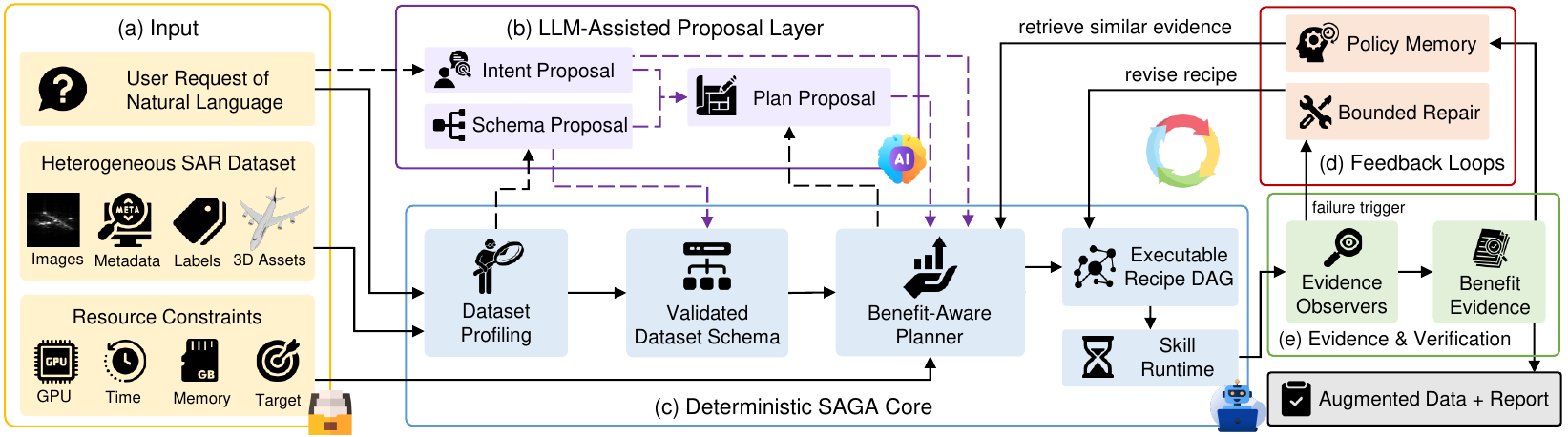}
    \caption{Overall architecture of SAGA. The system converts heterogeneous SAR inputs and natural-language requests into validated schemas, benefit-aware plans, executable recipe DAGs, and verified augmented datasets. Solid arrows denote deterministic execution and validated data flow, while dashed arrows denote LLM-assisted proposal flow that must be verified before execution. Feedback paths represent bounded intra-run repair and inter-run memory-based planning.}
    \label{fig:saga_overview}
\end{figure*}

This section presents the design of SAGA, a schema-grounded and benefit-aware agent framework for SAR data generation and augmentation.
Rather than treating augmentation as a single image-transformation or generation operation, SAGA formulates it as a data-centric decision problem. Given a heterogeneous user dataset, a natural-language request, available computational resources, and a library of augmentation skills, the system selects, compiles, executes, and verifies an augmentation recipe under explicit feasibility and evidence constraints.
The framework consists of six tightly coupled components: problem formulation and system overview, schema-grounded dataset profiling, benefit-aware hybrid planning, recipe-centric skill execution, observer-driven evaluation with bounded repair, and policy memory with evidence accumulation.

\subsection{Problem Formulation and System Overview}
\label{subsec:problem_overview}

Let $\mathcal{D}$ denote an input SAR dataset provided by the user.
The dataset may contain target chips, scene images, sidecar text files, captions, annotations, metadata embedded in filenames, or 3D assets.
Let $q$ denote the user's natural-language request, such as target-only augmentation, polarization-conditioned generation, sparse-azimuth completion, cross-platform feature migration, physical simulation, or target-background composition.
Let $\mathcal{C}$ denote resource and execution constraints, including GPU availability, runtime budget, memory limit, target sample count, and whether real execution or dry-run planning is required.
Let $\mathcal{K}=\{k_i\}_{i=1}^{N}$ be the SAR augmentation skill library, where each skill $k_i$ is associated with input requirements, controllable parameters, computational cost, expected benefits, and risks.
In this paper, a \emph{skill} denotes a domain-specific augmentation, generation, simulation, evaluation, or export capability with explicit input requirements, controllable parameters, expected outputs, cost profile, risk description, and standardized reporting interface.
Let $\mathcal{M}$ denote the policy memory that stores historical profiles, recipes, outcomes, and evaluation evidence.

SAGA maps these inputs to an augmented dataset, an executable recipe, and evidence reports:
\begin{equation}
\mathrm{SAGA}(\mathcal{D}, q, \mathcal{C}, \mathcal{K}, \mathcal{M})
\rightarrow
(\mathcal{D}_{aug}, \mathcal{R}, \mathcal{E}),
\label{eq:saga_mapping}
\end{equation}
where $\mathcal{D}_{aug}$ is the generated or augmented dataset, $\mathcal{R}$ is a reproducible recipe DAG, and $\mathcal{E}$ is a collection of observer and evaluator evidence.
The evidence may include quality checks, SAR artifact scores, metadata coverage, angle coverage, duplicate and leakage reports, distribution metrics, and optional downstream evaluation.

A central premise of SAGA is that SAR augmentation should not be reduced to optimizing a single image-generation objective.
Instead, SAGA selects and composes augmentation skills according to dataset deficits, user intent, execution constraints, and expected downstream utility.
For example, a speed-first request on a simple target-chip dataset may select a GAN-based generator, while sparse-angle aircraft extrapolation with an available 3D model may select GeoDiff-SAR, and cross-platform migration with reference images may select a style-transfer skill.
Accordingly, the central problem is not merely whether additional images can be produced, but whether a suitable, feasible, and verifiable augmentation strategy can be selected for the current SAR dataset and task.

Figure~\ref{fig:saga_overview} illustrates the overall architecture.
SAGA first profiles the dataset and grounds heterogeneous formats into a validated schema.
The intent recognizer and hybrid planner then combine the user request, dataset profile, resource constraints, skill descriptors, and policy memory to generate candidate recipes.
The selected recipe is executed by deterministic skill runtimes, observed by evaluators, optionally repaired within bounded limits, and finally exported with evidence reports.

\subsection{Schema-Grounded Dataset Profiling}
\label{subsec:schema_profiling}

SAR datasets are often organized in laboratory-specific or task-specific formats.
Important semantic fields, such as target class, azimuth angle, depression angle, resolution, band, polarization, sensor platform, or simulation parameters, may appear in directory names, filenames, sidecar text files, image captions, annotation files, or user-provided descriptions.
Requiring all users to convert data into a predetermined format would reduce practical usability, whereas allowing an LLM to freely infer and execute on unknown formats would introduce substantial reliability risks.
SAGA therefore adopts schema-grounded dataset profiling.

The profiler operates in three stages.
First, a deterministic raw scanner extracts observable facts from $\mathcal{D}$ without assuming a fixed dataset format.
These facts include file extensions, image counts, image sizes, bit depths, directory structures, sidecar files, annotation candidates, filename token patterns, numeric and string tokens, class-folder candidates, and image-label alignment statistics.
This stage does not assign semantic meaning to ambiguous tokens; it only records what can be objectively observed.

Second, SAGA induces a candidate dataset schema from the raw profile and the user's format hints.
For standard formats such as class-folder, YOLO, COCO, VOC, or mask-folder layouts, deterministic parsers can directly propose a schema.
For non-standard formats, an LLM-assisted schema induction module reads only the compact raw profile and user descriptions, rather than the entire dataset, and proposes candidate semantic mappings.
For example, if the user states that the tokens after ``850'' in a filename represent depression angle, azimuth angle, resolution, and band, the schema induction module maps these token positions to corresponding metadata fields.

Third, the candidate schema is passed through a deterministic validator.
The validator checks whether the proposed rules can be applied consistently to the dataset, whether the required fields are parseable, whether image-label pairs are aligned, whether metadata coverage is sufficient, and whether user-specified filters can select valid samples.
Only a schema that passes validation is accepted as the internal dataset schema.
In implementation, this validated schema is represented as validated dataset schema, but throughout the paper we refer to it as a validated dataset schema. This design explicitly separates semantic proposal from factual validation: LLM outputs and user hints propose candidate semantics, whereas programmatic validators verify their consistency against observable dataset facts.

Consequently, SAGA does not assume a fixed external dataset format, but it requires a deterministic internal schema before any skill is executed.
If the semantic interpretation is incomplete, SAGA produces a partial profile and requests the minimum missing information instead of executing on unsupported assumptions.

\begin{figure}[t]
    \centering
    \includegraphics[width=0.78\linewidth]{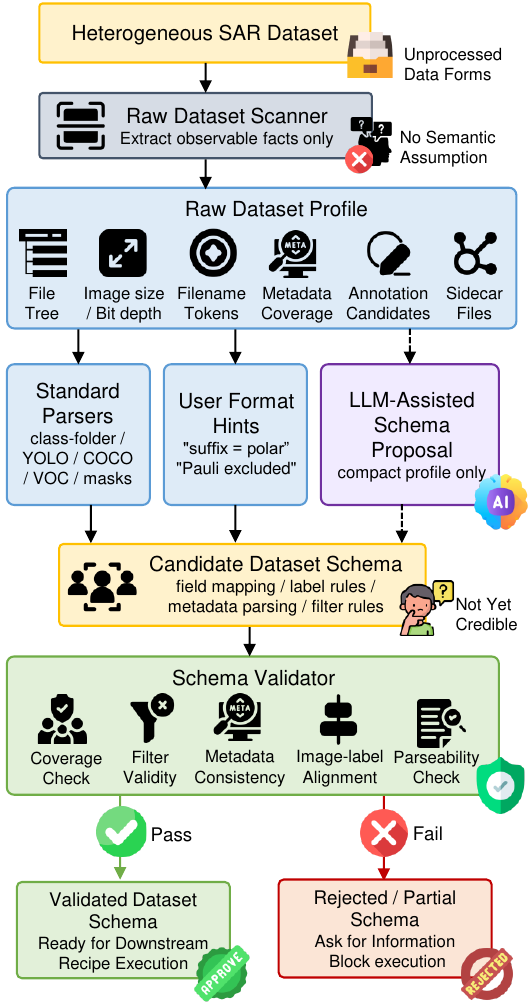}
    \caption{Schema-grounded dataset profiling. SAGA first extracts low-level observable facts from heterogeneous SAR datasets without semantic assumptions. Candidate schemas are then induced from standard parsers, user format hints, or optional LLM-assisted proposals. Only schemas that pass deterministic validation are accepted for downstream recipe execution; otherwise, SAGA rejects the schema or requests missing information.}
    \label{fig:schema_profiling}
\end{figure}

The complete schema-grounding workflow is illustrated in Figure~\ref{fig:schema_profiling}.

\subsection{Benefit-Aware Hybrid Planning}
\label{subsec:benefit_planning}

\begin{figure*}[t]
    \centering
    \includegraphics[width=0.95\textwidth]{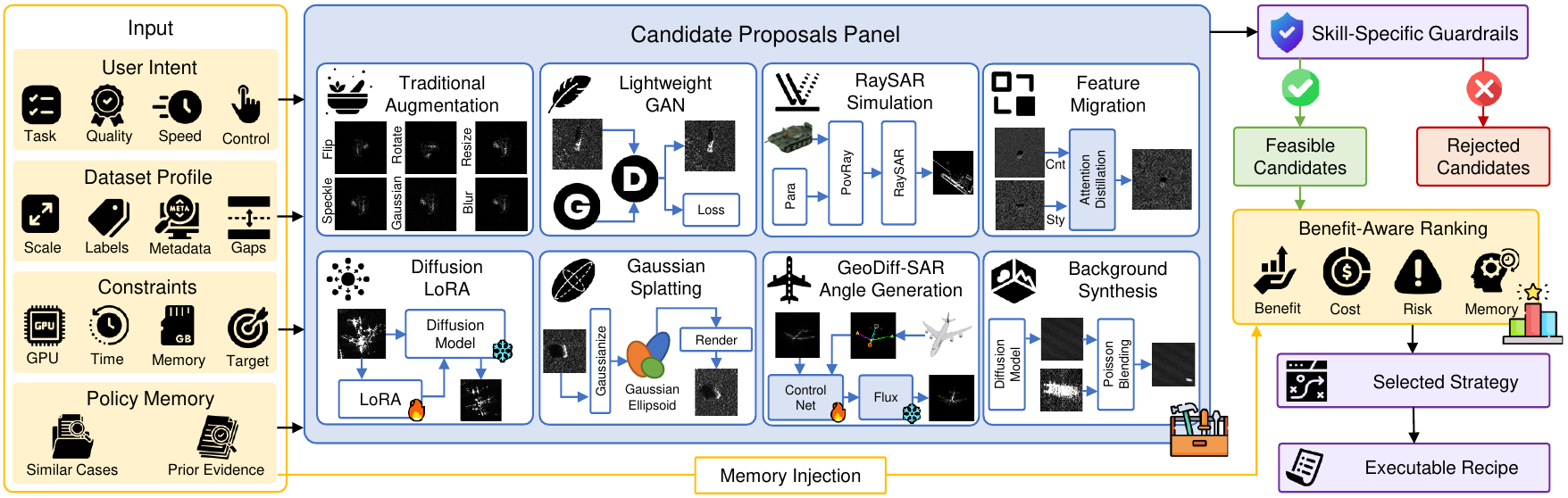}
    \caption{Benefit-aware hybrid planning. Candidate augmentation strategies are proposed from the user intent, dataset profile, execution constraints, and policy memory. They are verified by skill-specific guardrails, filtered according to input compatibility and resource feasibility, ranked using expected benefit, cost, risk, and memory evidence, and finally compiled into an executable recipe.}
    \label{fig:benefit_planning}
\end{figure*}

After schema grounding, SAGA determines which augmentation strategy is most appropriate for the current dataset, request, and resource constraints.
This decision is non-trivial because SAR augmentation skills differ substantially in capability, cost, controllability, and risk.
Traditional SAR-aware transformations are fast and robust but offer limited novelty.
GAN-based generation is efficient for simple target chips but has weak fine-grained controllability.
Diffusion LoRA generation offers higher quality and metadata-conditioned synthesis but requires training time and GPU memory.
GeoDiff-SAR supports geometry-guided large-angle extrapolation but is computationally heavy and requires 3D priors.
RaySAR provides physically interpretable simulation from 3D models and imaging parameters, but may exhibit domain gaps.
Style transfer supports cross-platform feature migration, while target-background composition supports scene-level synthesis.

SAGA uses a hybrid planner in which an LLM provides high-level intent understanding and candidate plan proposals, while deterministic rules verify feasibility and rank candidate plans.
The planner receives the validated dataset profile $\phi(\mathcal{D})$, the structured user intent $\psi(q)$, the resource constraints $\mathcal{C}$, skill descriptors $\mathcal{K}$, and relevant memory evidence from $\mathcal{M}$.
For each candidate skill or recipe $r_i$, SAGA estimates a planning score:
\begin{equation}
\begin{aligned}
\mathrm{score}(r_i)
=&\ \alpha B\!\left(r_i;\phi(\mathcal{D}),\psi(q)\right) - \beta C\!\left(r_i;\mathcal{C}\right) \\
&- \gamma R\!\left(r_i;\phi(\mathcal{D})\right) + \delta M\!\left(r_i;\mathcal{M}\right).
\end{aligned}
\label{eq:planning_score}
\end{equation}
where $B$ is the expected dataset or task benefit, $C$ is the execution cost, $R$ is the risk of invalid or harmful augmentation, and $M$ is the support from historical evidence.
The coefficients control the relative importance of benefit, cost, risk, and memory evidence.
In practice, the score is implemented as a transparent rule-based ranking function informed by LLM-parsed intent and skill capability descriptors.

The planner is constrained by guardrails.
A candidate skill is rejected or down-weighted when its input requirements are not satisfied, when required metadata is unavailable, when the resource budget is insufficient, or when the requested task is better addressed by another skill.
For example, RaySAR requires a 3D model and radar imaging parameters; GeoDiff-SAR requires sparse real images and geometry/physical priors; diffusion LoRA benefits from captions or metadata that can be converted into captions; style transfer requires both content and style domains.
This constraint prevents the LLM from directly invoking tools in an unconstrained ReAct-style manner.

The output of the planner is not an executed tool call, but a structured plan containing the selected task, selected skill, rejected alternatives, selection reasons, resource settings, expected outputs, and required evaluators.
This plan is then compiled into an executable recipe DAG.
Therefore, SAGA is neither a keyword-based router nor a free-form LLM tool caller.
It is a benefit-aware hybrid planner: the LLM proposes high-level semantic interpretations, while deterministic guardrails enforce feasibility and the ranker selects a verifiable augmentation strategy.

The benefit-aware hybrid planning process is illustrated in Figure~\ref{fig:benefit_planning}.

\subsection{Recipe-Centric Skill Execution}
\label{subsec:recipe_execution}

To support reproducibility and auditability, SAGA does not directly execute LLM-generated tool calls.
Instead, the selected plan is compiled into an executable recipe DAG:
\begin{equation}
\mathcal{R}=(V,E),
\end{equation}
where $V$ is a set of nodes and $E$ is a set of dependency edges.
Each node $v_i \in V$ corresponds to a profiling step, preprocessing step, augmentation skill, evaluator, repair policy, or export operation.
We denote a recipe node as:
\begin{equation}
v_i = (k_i, x_i, \theta_i, y_i, \rho_i),
\end{equation}
where $k_i$ is the skill or operation, $x_i$ is the input artifact binding, $\theta_i$ is the parameter set, $y_i$ is the expected output artifact, and $\rho_i$ is the standardized run report.

The recipe abstraction provides several advantages.
First, it makes the execution path explicit: each augmentation run records which dataset schema, filters, skill parameters, model paths, random seeds, and evaluation steps were used.
Second, it enables dry-run and real-run separation.
In dry-run mode, SAGA validates inputs and writes the recipe, commands, and expected artifacts without expensive model execution.
In real-run mode, the same recipe can be executed to materialize outputs.
Third, it allows failed or partially completed runs to be resumed or analyzed step by step.
Fourth, it provides a common interface for heterogeneous skills, including traditional augmentation, GAN generation, diffusion LoRA, GeoDiff-SAR, RaySAR simulation, style transfer, background generation, target-background composition, and evaluation skills.

Each executed skill writes a standardized standardized execution report, including status, inputs, outputs, parameters, commands, elapsed time, generated count, artifact paths, warnings, and failure messages.
This report becomes part of the provenance record and is consumed by observers, exporters, and memory modules.
By converting high-level plans into recipe DAGs and standardized run reports, SAGA turns agentic augmentation from ad hoc tool invocation into reproducible workflow execution.

\begin{figure}[t]
    \centering
    \includegraphics[width=0.9\linewidth]{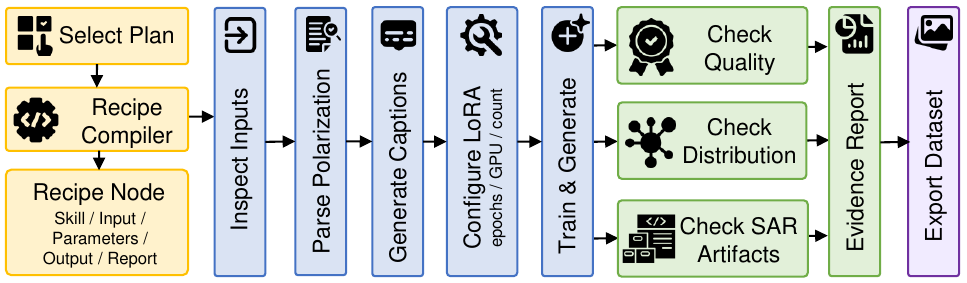}
    \caption{Recipe-centric skill execution. Planner outputs are compiled into executable recipe DAGs rather than directly executed as tool calls. Each node binds a skill or operation with input artifacts, parameters, expected outputs, and standardized run reports. The example recipe illustrates polarization-conditioned LoRA generation followed by quality, distribution, and SAR-artifact evaluations.}
    \label{fig:recipe_execution}
\end{figure}

Figure~\ref{fig:recipe_execution} visualizes how a selected plan is compiled into an executable recipe DAG.

\subsection{Observer-Driven Evaluation and Bounded Repair}
\label{subsec:observer_repair}

SAGA explicitly distinguishes file-level generation success from augmentation utility.
A generated image may exist on disk but remain unsuitable for training because of invalid metadata, label mismatch, excessive artifacts, near-duplicate content, data leakage, poor diversity, or failure to address the targeted dataset deficit.
Therefore, SAGA attaches observer evidence to each produced dataset.

\begin{figure*}[t]
    \centering
    \includegraphics[width=0.95\textwidth]{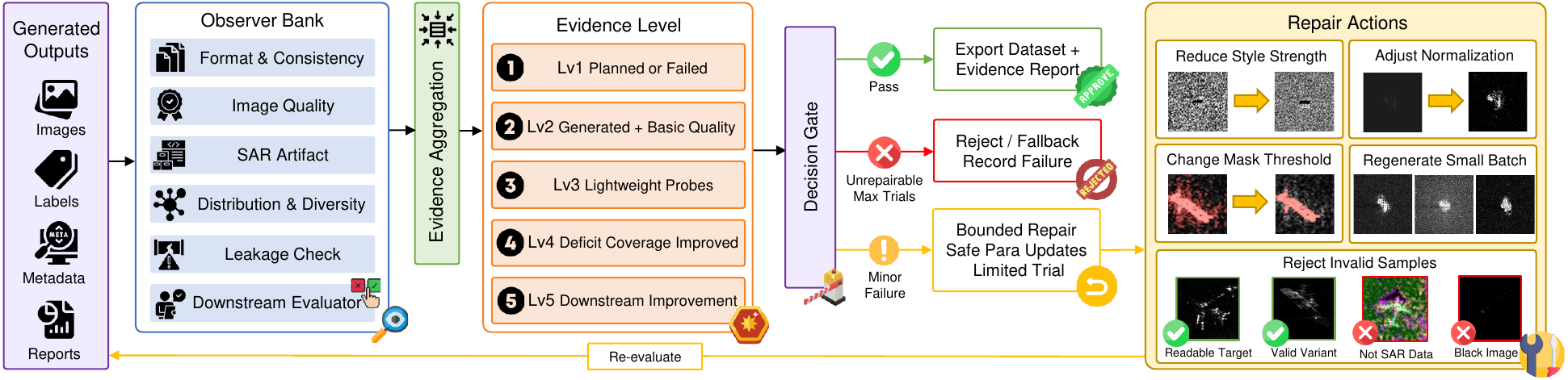}
    \caption{Observer-driven evaluation and bounded repair. Generated outputs are assessed by observers covering format consistency, image quality, SAR artifacts, distribution and diversity, leakage, and optional downstream evaluators. Observer results are aggregated into evidence levels that distinguish generation success from augmentation benefit. Repair is triggered only by explicit observer failures and is restricted to safe parameter updates under a bounded number of trials.}
    \label{fig:observer_repair}
\end{figure*}

The observer layer consists of several evaluator groups.
Format and consistency observers check file readability, image-label alignment, metadata preservation, expected output count, and valid export structure.
Image quality observers check dynamic range, black/white image ratio, corrupted files, and basic statistics.
SAR artifact observers examine SAR-specific failure modes such as stripe artifacts, abnormal smooth background gradients, target fragmentation, target compactness, and centering when appropriate.
Distribution and diversity observers compare generated and reference images using lightweight feature statistics and duplicate or near-duplicate checks.
Leakage observers detect whether augmented samples are too close to validation or test data.
When the user requests a downstream claim, optional task evaluators such as classification probes can be used.

To avoid overclaiming augmentation benefit, SAGA assigns evidence levels to the generated outputs:
\begin{itemize}
    \item \textbf{Level 1}: a recipe is planned, failed, or only dry-run evidence is available;
    \item \textbf{Level 2}: samples are generated and basic quality checks pass;
    \item \textbf{Level 3}: lightweight distribution, diversity, metadata, or artifact probes pass;
    \item \textbf{Level 4}: a dataset deficit is measurably improved, such as class balance, polarization coverage, or angle coverage;
    \item \textbf{Level 5}: a downstream evaluator reports task-level improvement.
\end{itemize}
Only Level~5 evidence supports a downstream performance-gain claim.
Lower evidence levels provide increasingly strong, but still indirect, evidence that the generated samples are usable.

When observer triggers are activated, SAGA enters a bounded repair loop.
The repair loop is not an open-ended LLM self-reflection process.
It is constrained by three rules.
First, repair is triggered only by explicit observer failures.
Second, the number of repair trials is bounded by a maximum value.
Third, each repair action can modify only a predefined safe parameter set, such as normalization mode, style strength, mask threshold, feather radius, generation count, sample rejection threshold, or prompt selection.
If repair fails, SAGA records the failure and falls back to the last stable artifact or rejects invalid samples.
This design improves robustness while preserving reproducibility.

\begin{figure}[t]
    \centering
    \includegraphics[width=0.9\linewidth]{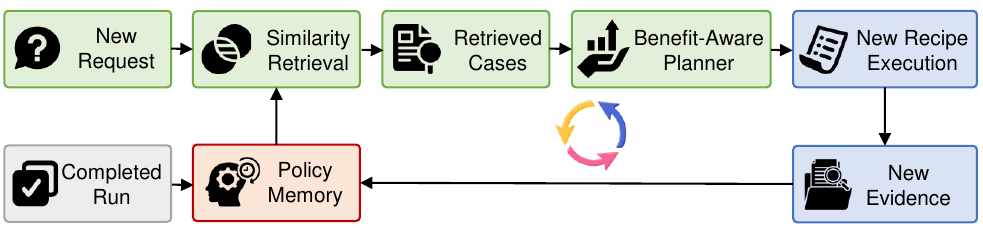}
    \caption{Policy memory and evidence accumulation. SAGA records dataset profiles, recipes, outcomes, and evidence levels as structured case memories. For future requests, similar historical cases are retrieved as planning evidence, providing success patterns, failure warnings, and parameter hints for benefit-aware augmentation planning.}
    \label{fig:policy_memory}
\end{figure}

The observer-driven evaluation and bounded repair mechanism is summarized in Figure~\ref{fig:observer_repair}.

\subsection{Policy Memory and Evidence Accumulation}
\label{subsec:policy_memory}

The final component of SAGA is an evidence-based policy memory.
The purpose of memory is not to claim that SAGA has learned an optimal augmentation policy.
Instead, memory records previous runs and uses them as planning evidence for future similar datasets and requests.
Each memory entry stores a dataset profile, structured intent, selected recipe, skill configuration, execution outcome, observer results, and evidence level:
\begin{equation}
m_j =
(\phi(\mathcal{D}_j), \psi(q_j), \mathcal{R}_j, o_j, e_j),
\end{equation}
where $o_j$ denotes execution outcomes and $e_j$ denotes evaluation evidence.

During planning, SAGA retrieves memory entries with similar dataset profiles, tasks, metadata deficits, or resource constraints.
These entries provide positive or negative evidence for candidate skills.
For instance, previous runs may show that GAN generation is effective for simple target chips under limited memory, that diffusion LoRA benefits from metadata-derived captions, that GeoDiff-SAR is preferable for large-angle aircraft extrapolation when a 3D model is available, or that style transfer is more suitable than retraining when cross-platform reference images exist.
Conversely, failed runs can down-weight skills under unsupported data conditions.

Memory evidence is used only as a planning signal and remains subject to the same schema validation, skill guardrails, and observer checks as new requests.
Thus, SAGA accumulates reusable evidence without bypassing deterministic verification.
Over time, this memory allows the system to move from isolated tool execution toward evidence-informed SAR augmentation planning.

Figure~\ref{fig:policy_memory} shows how SAGA stores and retrieves evidence-based policy memory across runs.

Overall, SAGA combines schema-grounded profiling, benefit-aware hybrid planning, recipe-centric execution, observer-driven verification, bounded repair, and evidence-based memory.
This design allows the system to handle heterogeneous SAR datasets and diverse augmentation skills while maintaining reproducibility, controllability, and cautious benefit claims.
 
\section{Experiments}

\subsection{Experimental Protocol}

We evaluate SAGA along two complementary axes: \emph{agentic reliability} and \emph{augmentation utility}. 
Agentic reliability measures whether SAGA can ground heterogeneous SAR datasets, interpret natural-language augmentation requests, select feasible skills, compile executable recipes, and validate generated outputs. 
Augmentation utility measures whether the selected augmentation strategies improve downstream SAR interpretation tasks.

Because SAGA is not a single SAR image generator, evaluating it solely by image-quality metrics against GAN- or diffusion-based models would be insufficient. We therefore evaluate whether SAGA makes reliable data-conditioned augmentation decisions and whether these decisions translate into evidence-qualified downstream benefits. The experiments are organized as follows:
\begin{itemize}
    \item Exp.1 evaluates schema-grounded dataset profiling on heterogeneous SAR dataset layouts.
    \item Exp.2 evaluates natural-language intent recognition and guarded skill planning.
    \item Exp.3 evaluates whether high-level plans can be compiled into reproducible recipe DAGs.
    \item Exp.4 evaluates observer-driven validation and bounded repair under controlled failure injection.
    \item Exp.5 evaluates downstream augmentation benefit across multiple SAR task groups.
    \item Exp.6 provides qualitative case studies and traceability examples.
    \item Exp.7 performs ablation analysis over the main SAGA modules.
\end{itemize}

For controlled agentic benchmarks, including Exp.1, Exp.2, and Exp.4, the reported cases are expert-defined stress tests designed to isolate known failure modes. These results should therefore be interpreted as controlled reliability measurements rather than open-world user success rates. For small controlled benchmarks, we report raw counts in tables and use Jeffreys-smoothed rates in figures:
\begin{equation}
\hat{p}=\frac{s+0.5}{n+1},
\end{equation}
where $s$ denotes the number of successful cases and $n$ denotes the total number of cases.

We use the evidence levels defined in Sec.~\ref{sec:method}. In particular, Lv5 evidence is assigned only when the generated training set passes observer and leakage gates, a downstream evaluator is executed, and the improvement over the no-augmentation baseline exceeds a predefined practical margin or passes a paired significance test. Therefore, only Lv5 supports a downstream performance gain claim.

\begin{figure}[t]
    \centering
    \includegraphics[width=0.9\linewidth]{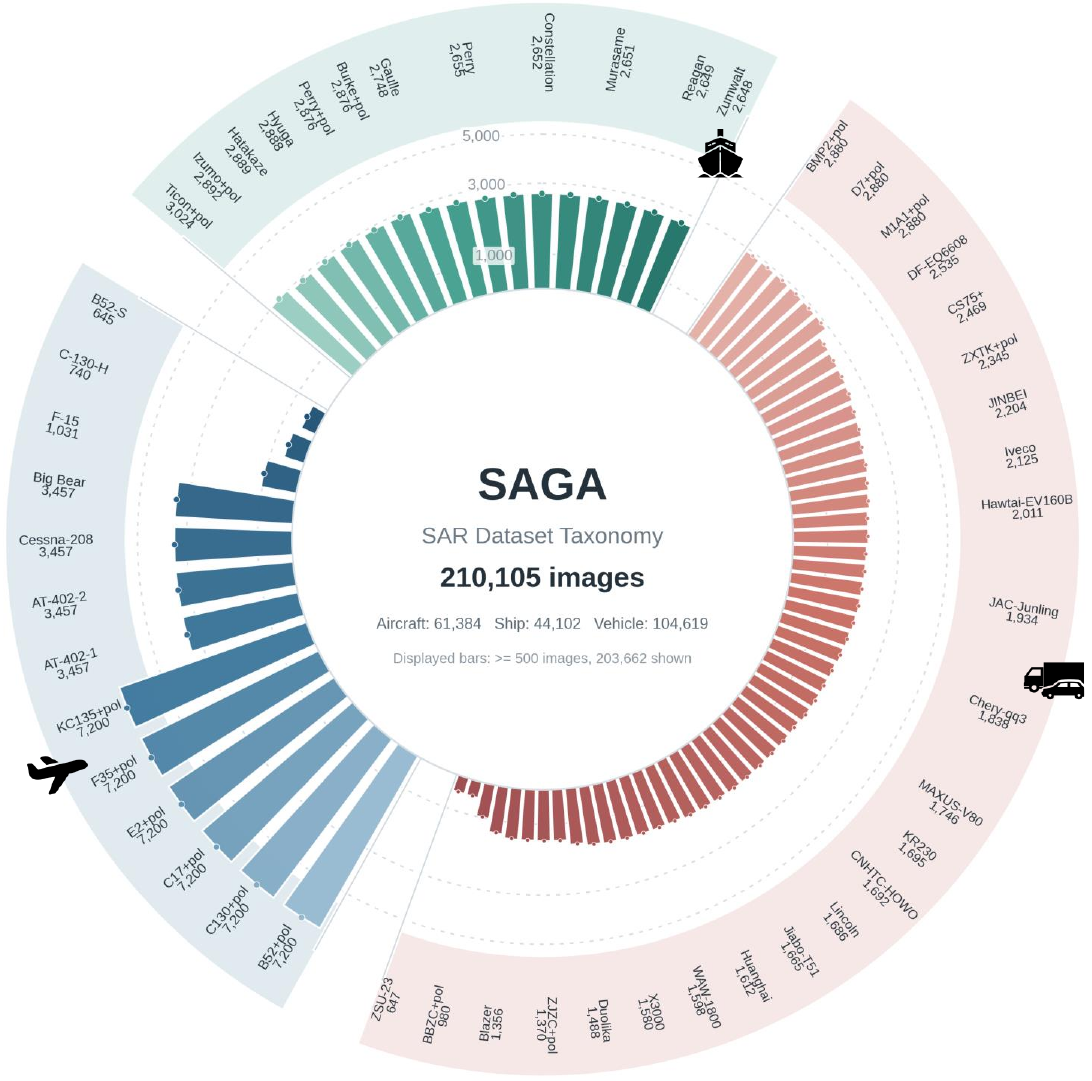}
    \caption{SAR dataset taxonomy used in the SAGA evaluation. The dataset contains 210,105 images across aircraft, ship, and vehicle categories. Displayed bars correspond to classes with at least 500 images, illustrating the heterogeneous and long-tailed class distribution considered in the experimental setup.}
    \label{fig:dataset_taxonomy}
\end{figure}

Figure~\ref{fig:dataset_taxonomy} summarizes the SAR dataset taxonomy used to organize the experimental resources. The collection contains 210,105 SAR images from aircraft, ship, and vehicle categories, including 61,384 aircraft images, 44,102 ship images, and 104,619 vehicle images. The figure highlights the long-tailed class distribution and the diversity of target categories, which motivate the need for schema-grounded profiling and benefit-aware augmentation planning.

\subsection{Dataset Profiling and Schema Grounding}

\subsubsection{Setup}

Exp.1 evaluates whether SAGA can infer executable dataset schemas from heterogeneous SAR data organizations. The benchmark contains 35 controlled schema cases covering class-folder target chips, filename metadata, nested directory metadata, key-value sidecar files, caption sidecars, dataset-level constant fields, mixed path-and-filename metadata, multi-class filtering, and several invalid or incomplete requests such as missing polarization, missing incidence angle, unsupported azimuth filtering, and conflicting user hints.

We compare five variants: a fallback parser with validator, a rule-based parser with validator, a user-hint parser with validator, Full SAGA without validator, and Full SAGA. The main metrics are schema decision accuracy, valid-case acceptance, invalid-case rejection, required-field coverage, metadata value accuracy, and invalid acceptance rate.

\begin{figure}[t]
    \centering
    \includegraphics[width=0.78\linewidth]{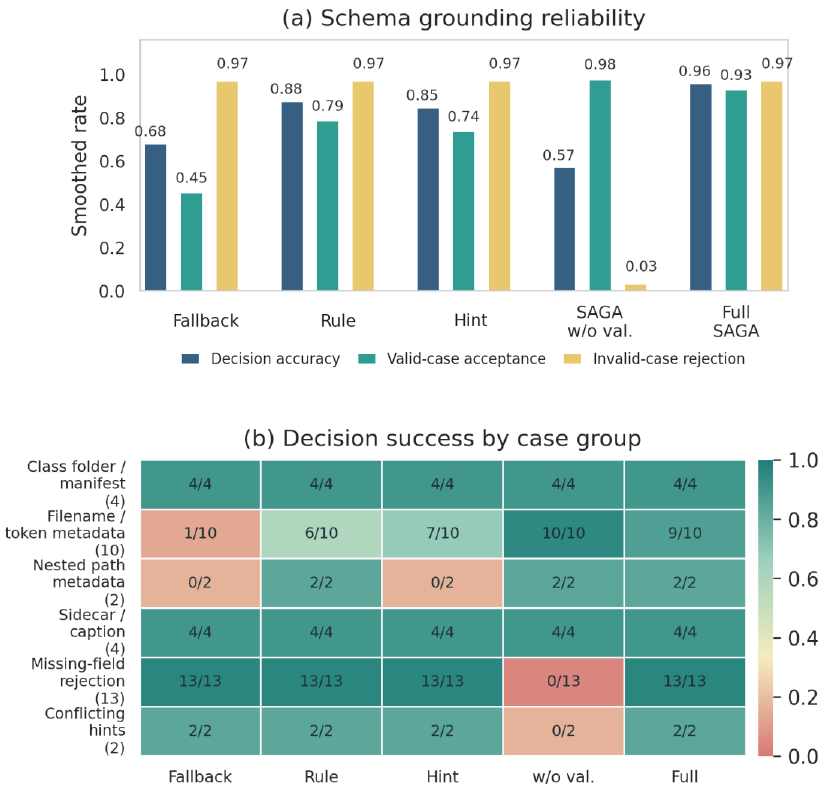}
    \caption{Dataset profiling and schema-grounding performance. SAGA first extracts observable dataset facts and then accepts only validator-gated schemas. The case-group heatmap highlights that the no-validator variant fails on missing-field and conflicting-hint cases, whereas Full SAGA rejects these invalid schema interpretations.}
    \label{fig:exp1_schema}
\end{figure}

\begin{table}[t]
\centering
\caption{Dataset profiling and schema-grounding results on 35 controlled schema cases. Full SAGA achieves 34/35 correct schema decisions, accepts 19/20 valid schemas, and rejects 15/15 invalid schemas.}
\label{tab:exp1_schema}
{\scriptsize
\setlength{\tabcolsep}{2.0pt}
\renewcommand{\arraystretch}{1.08}
\begin{tabular*}{\columnwidth}{@{\extracolsep{\fill}}lcccccc@{}}
\hline
Method & Dec. & Valid & Rej. & Field & Value & Inv. \\
\hline
Fallback & 68.6 & 45.0 & 100.0 & 53.4 & 62.3 & 0.0 \\
Rule & 88.6 & 80.0 & 100.0 & 80.2 & 93.6 & 0.0 \\
Hint & 85.7 & 75.0 & 100.0 & 70.0 & 80.2 & 0.0 \\
w/o Val. & 57.1 & 100.0 & 0.0 & 85.3 & 99.3 & 100.0 \\
Full SAGA & \textbf{97.1} & \textbf{95.0} & \textbf{100.0} & \textbf{85.3} & \textbf{99.3} & \textbf{0.0} \\
\hline
\end{tabular*}}
\vspace{0.5mm}
\begin{flushleft}
\footnotesize
Dec.: schema decision accuracy; Valid: valid-case acceptance; Rej.: invalid-case rejection; Inv.: invalid-schema acceptance.
\end{flushleft}
\end{table}

\subsubsection{Results}

As shown in Fig.~\ref{fig:exp1_schema} and Table~\ref{tab:exp1_schema}, Full SAGA achieves the best overall schema decision accuracy and rejects all invalid schemas. The no-validator variant achieves high field coverage and metadata value accuracy because it aggressively accepts semantic proposals, but it also accepts all invalid schemas. This result supports the central design choice of SAGA: LLM- or user-proposed schemas are treated as verifiable hypotheses rather than executable facts.

Full SAGA accepts most valid cases, rejects all invalid cases, and maintains high metadata value accuracy. Its only schema-decision failure occurs in a valid but difficult mixed-format case, indicating that the benchmark is not solved by trivial pattern matching. Overall, the result supports the schema-grounding design: SAGA first performs observable raw profiling, then induces candidate semantic proposals, and finally accepts only deterministic validator-gated schemas for downstream execution.

\subsection{Intent Recognition and Skill Planning}

\subsubsection{Setup}

Exp.2 evaluates whether SAGA can map natural-language requests and dataset profiles to compatible augmentation strategies. The benchmark contains 80 controlled planning cases across 11 categories: explicit generation, metadata-conditioned generation, geometry-conditioned generation, implicit benefit planning, physics simulation, multi-input composition, explicit transformation, negative guardrail cases, downstream-evidence requests, hard argument binding, and hard multi-step clarification.

We compare Full SAGA with four baselines: Keyword Router, Rule-only Planner, LLM-only Planner, and ReAct-style Agent. The metrics include Top-1 and Top-3 skill accuracy, executable-or-reject decision accuracy, invalid selection rate, observer recall, argument accuracy, rejected-skill accuracy, recipe skeleton success, and failure-free rate. Top-1, Top-3, decision, recipe, and failure-free metrics are case-level over 80 planning cases. Invalid selection is computed over 19 infeasible-skill cases. Observer recall, argument accuracy, and rejected-skill accuracy are computed over the corresponding annotated labels and slots.

\begin{figure*}[t]
    \centering
    \includegraphics[width=0.95\textwidth]{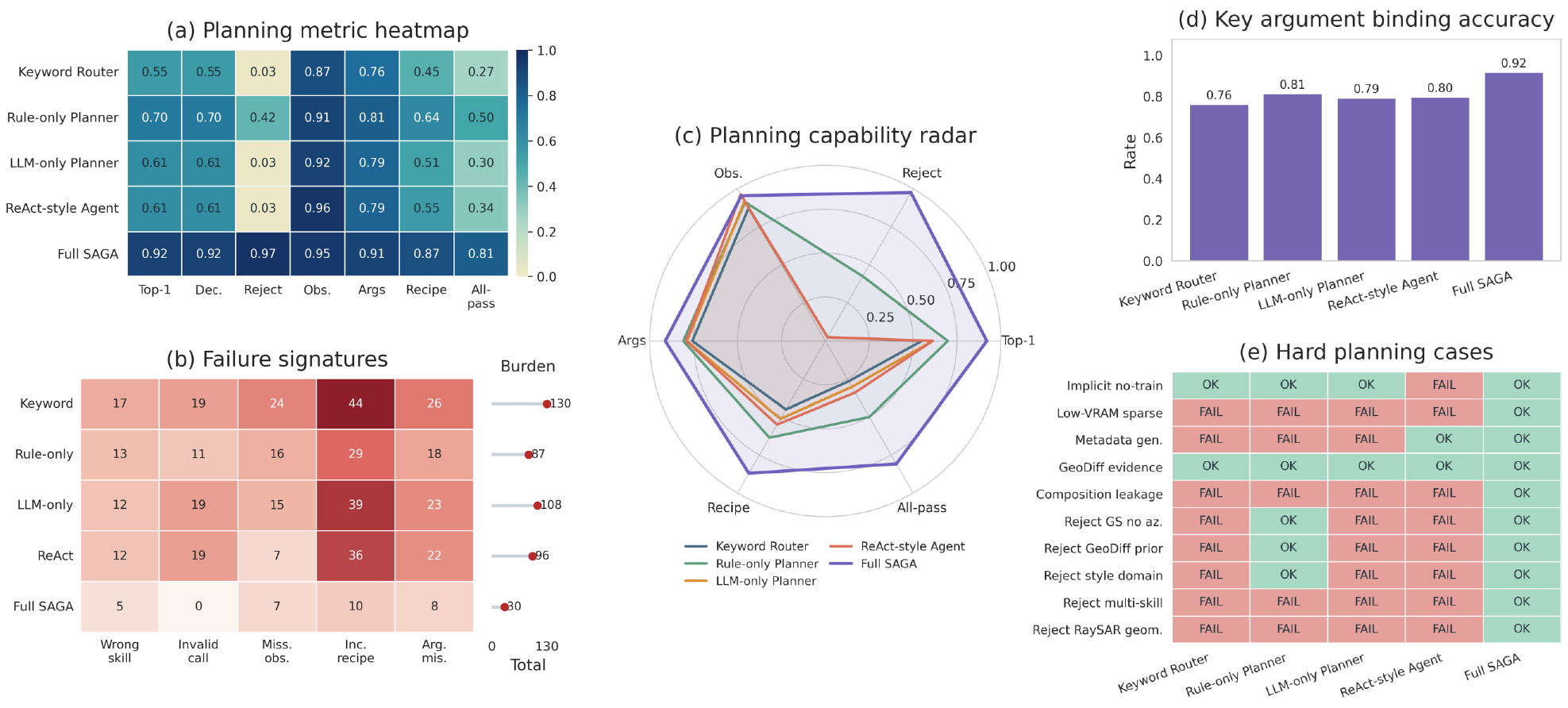}
    \caption{Intent recognition and skill planning results. SAGA improves both skill selection and invalid-call avoidance. Compared with LLM-only and ReAct-style baselines, SAGA is less likely to invoke infeasible tools because skill proposals are verified by dataset schema and skill-specific guardrails.}
    \label{fig:exp2_planning}
\end{figure*}

\begin{table}[t]
\centering
\caption{Planning performance on 80 controlled natural-language augmentation requests. Full SAGA makes 74/80 correct planning decisions, avoids all 19 infeasible skill calls, and completes 65/80 cases without planning failures.}
\label{tab:exp2_planning}
{\scriptsize
\setlength{\tabcolsep}{1.6pt}
\renewcommand{\arraystretch}{1.08}
\begin{tabular*}{\columnwidth}{@{\extracolsep{\fill}}lccccccccc@{}}
\hline
Method & T1 & T3 & Dec. & Inv. & Obs. & Arg. & Rej. & Rec. & FF \\
\hline
Key & 55.0 & 55.0 & 55.0 & 100.0 & 87.8 & 76.0 & 0.0 & 45.0 & 26.2 \\
Rule & 70.0 & 70.0 & 70.0 & 57.9 & 91.8 & 81.2 & 7.9 & 63.7 & 50.0 \\
LLM & 61.2 & 66.2 & 61.2 & 100.0 & 92.4 & 79.2 & 0.0 & 51.2 & 30.0 \\
ReAct & 61.2 & 71.2 & 61.2 & 100.0 & 96.9 & 79.6 & 0.0 & 55.0 & 33.8 \\
SAGA & \textbf{92.5} & \textbf{97.5} & \textbf{92.5} & \textbf{0.0} & \textbf{96.0} & \textbf{91.7} & \textbf{100.0} & \textbf{87.5} & \textbf{81.2} \\
\hline
\end{tabular*}}
\vspace{0.5mm}
\begin{flushleft}
\footnotesize
T1/T3: Top-1/Top-3 skill accuracy; Inv.: invalid selection rate; Rec.: recipe skeleton success; FF: failure-free rate.
\end{flushleft}
\end{table}

\subsubsection{Results}

Fig.~\ref{fig:exp2_planning} and Table~\ref{tab:exp2_planning} show that Full SAGA substantially improves planning reliability under the controlled benchmark protocol. It achieves 92.5\% Top-1 skill accuracy and 97.5\% Top-3 skill accuracy. More importantly, it avoids all infeasible skill calls on the 19 negative guardrail cases. In contrast, LLM-only and ReAct-style baselines often attach plausible observers or produce fluent plans, but they still invoke incompatible skills when required metadata, physical priors, style domains, or 3D assets are missing.

Despite these gains, Full SAGA is not failure-free. Its failure-free rate is 81.2\%, with remaining errors mainly coming from hard cases involving implicit strategy selection, missing optional observers, incomplete recipe skeletons, and numeric argument binding. This result indicates that the benchmark is not saturated and that the remaining errors mainly arise from genuinely ambiguous planning cases.

\subsection{Recipe Execution Reliability}

\subsubsection{Setup}

Exp.3 evaluates whether high-level plans can be compiled into executable and reproducible recipe DAGs. We use nine representative execution cases, including real runs for lightweight skills such as traditional augmentation, pseudocolor conversion, and target-background composition, and dry-run validation for heavier or dependency-sensitive skills such as LoRA generation, GAN generation, Gaussian splatting, GeoDiff-SAR, and RaySAR sweep. Dry-run validation checks recipe legality, artifact binding, parameter binding, and report generation without executing the expensive model.

\begin{table}[t]
\centering
\caption{Recipe execution reliability. Protocol Pass denotes whether mandatory compile, execution, output, replay, observer/export, and report requirements are satisfied.}
\label{tab:exp3_recipe}
{\scriptsize
\setlength{\tabcolsep}{1.8pt}
\renewcommand{\arraystretch}{1.08}
\begin{tabular*}{\columnwidth}{@{\extracolsep{\fill}}lccccccccc@{}}
\hline
Method & Comp. & Exec. & Dry & Real & Out. & Rep. & Obs. & Replay & Pass \\
\hline
Script & 100.0 & 100.0 & 100.0 & 100.0 & 88.9 & 68.9 & 0.0 & 100.0 & 0.0 \\
LLM & 100.0 & 88.9 & 100.0 & 66.7 & 77.8 & 67.8 & 0.0 & 88.9 & 0.0 \\
SAGA & \textbf{100.0} & \textbf{100.0} & \textbf{100.0} & \textbf{100.0} & \textbf{100.0} & \textbf{96.7} & \textbf{100.0} & \textbf{100.0} & \textbf{100.0} \\
\hline
\end{tabular*}}
\end{table}

\subsubsection{Results}

As shown in Table~\ref{tab:exp3_recipe}, direct scripts and LLM-generated tool calls can execute some skills successfully, but they do not satisfy the SAGA protocol because they lack standardized reports, observer/export coverage, and complete provenance. A failed protocol pass does not necessarily mean that the script produced no files; it means that at least one mandatory SAGA reproducibility or evidence requirement was missing.

The report completeness of Full SAGA is 96.7\%, whereas its protocol pass rate is 100\%. These two metrics are not contradictory: the protocol pass rate measures whether all mandatory execution and provenance requirements are satisfied, while report completeness is a finer-grained field-level score that also includes optional metadata fields.

\subsection{Observer Evidence and Bounded Repair}

\subsubsection{Setup}

Exp.4 evaluates whether SAGA can detect invalid SAR augmentations and perform bounded repair. The benchmark includes 10 failure or stress batches and one clean reference batch. Failure types include black-heavy images, low dynamic range, stripe artifacts, abnormal background gradients, fragmented targets, off-center targets, near duplicates, validation/test leakage, mixed failures, and count/manifest consistency errors.

We compare four settings: No Observer, Quality Only, Full Observers, and Full Observers with bounded repair. Full Observers include quality checks, SAR artifact checks, duplicate and leakage checks, and count/metadata consistency checks. Bounded repair is triggered only by explicit observer failures and is restricted to safe parameter changes, invalid-sample removal, or manifest correction.

\begin{figure}[htbp]
    \centering
    \includegraphics[width=0.9\linewidth]{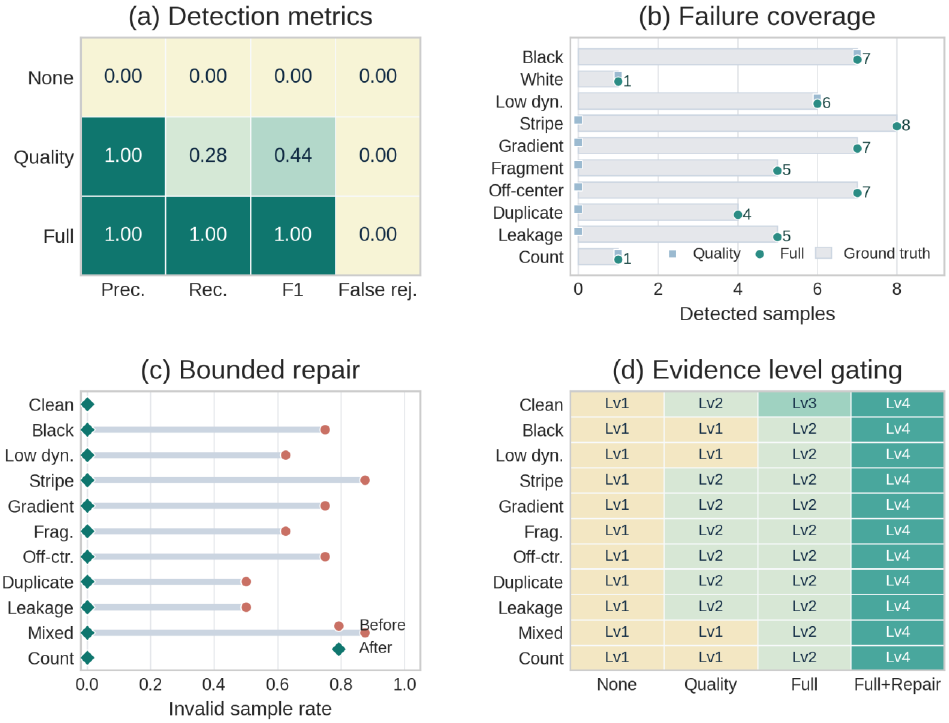}
    \caption{Observer-driven evaluation and bounded repair. Quality-only checks detect simple failures but miss SAR-specific artifacts, duplication, leakage, and consistency errors. Full observers improve detection coverage, while bounded repair reduces invalid exports but does not repair all failure modes. After bounded repair, only 6/11 batches reach Lv4 evidence, indicating that repair reduces invalid exports but does not eliminate all failures.}
    \label{fig:exp4_observer}
\end{figure}

\begin{table}[t]
\centering
\caption{Observer detection results under controlled failure injection. Full observers are not treated as an oracle; residual false rejection and missed failures remain.}
\label{tab:exp4_detection}
{\scriptsize
\setlength{\tabcolsep}{3.0pt}
\renewcommand{\arraystretch}{1.08}
\begin{tabular*}{\columnwidth}{@{\extracolsep{\fill}}lcccc@{}}
\hline
Method & Prec. & Rec. & F1 & False Rej. \\
\hline
No Observer & 0.0 & 0.0 & 0.0 & 0.0 \\
Quality Only & 100.0 & 28.0 & 43.8 & 0.0 \\
Full Observers & \textbf{93.8} & \textbf{90.0} & \textbf{91.8} & 8.6 \\
\hline
\end{tabular*}}
\end{table}

\subsubsection{Results}

Fig.~\ref{fig:exp4_observer} and Table~\ref{tab:exp4_detection} show that quality-only checks are insufficient for SAR augmentation validation. They detect simple failures such as black-heavy images and low dynamic range, but miss many SAR-specific problems. Full Observers improve recall to 90.0\% and F1 to 91.8\%, but they still produce an 8.6\% false-rejection rate, indicating that observer-based validation is not an oracle.

Bounded repair reduces invalid outputs for most predefined failure types. However, stripe artifacts, abnormal gradients, fragmented targets, leakage, and mixed failures still retain residual invalid rates after repair. Consequently, Full + Repair raises only 6 out of 11 batches to Lv4 evidence, rather than claiming universal repair. This result supports the SAGA evidence principle: file generation alone is not sufficient; generated data are exported only with evidence-level qualifications.

\subsection{Downstream Augmentation Benefit}

\subsubsection{Setup}

Exp.5 evaluates whether SAGA-selected augmentation strategies improve downstream SAR tasks. We use eight task groups: low-shot vehicle ATR, imbalanced full-polarization vehicles, cross-polarization ATR, cross-angle vehicle ATR, low-label ship detection, target-background scenes, sparse-view aircraft, and sim-to-real vehicle ATR. Each task-method setting is evaluated with 12 seeds, resulting in 96 matched runs per method.

We compare No Augmentation, Traditional augmentation, Fixed GAN, Fixed Diffusion, Fixed Polarization, Fixed Simulation, Rule-only Selection, LLM-only Selection, Manual Expert Policy, SAGA without observer, and Full SAGA. Score and gain are reported as mean $\pm$ standard deviation over matched task-seed pairs. Cost is a normalized augmentation cost index computed from skill runtime class and task resource factor; downstream training and evaluator costs are excluded.

\begin{figure*}[t]
    \centering
    \includegraphics[width=0.95\textwidth]{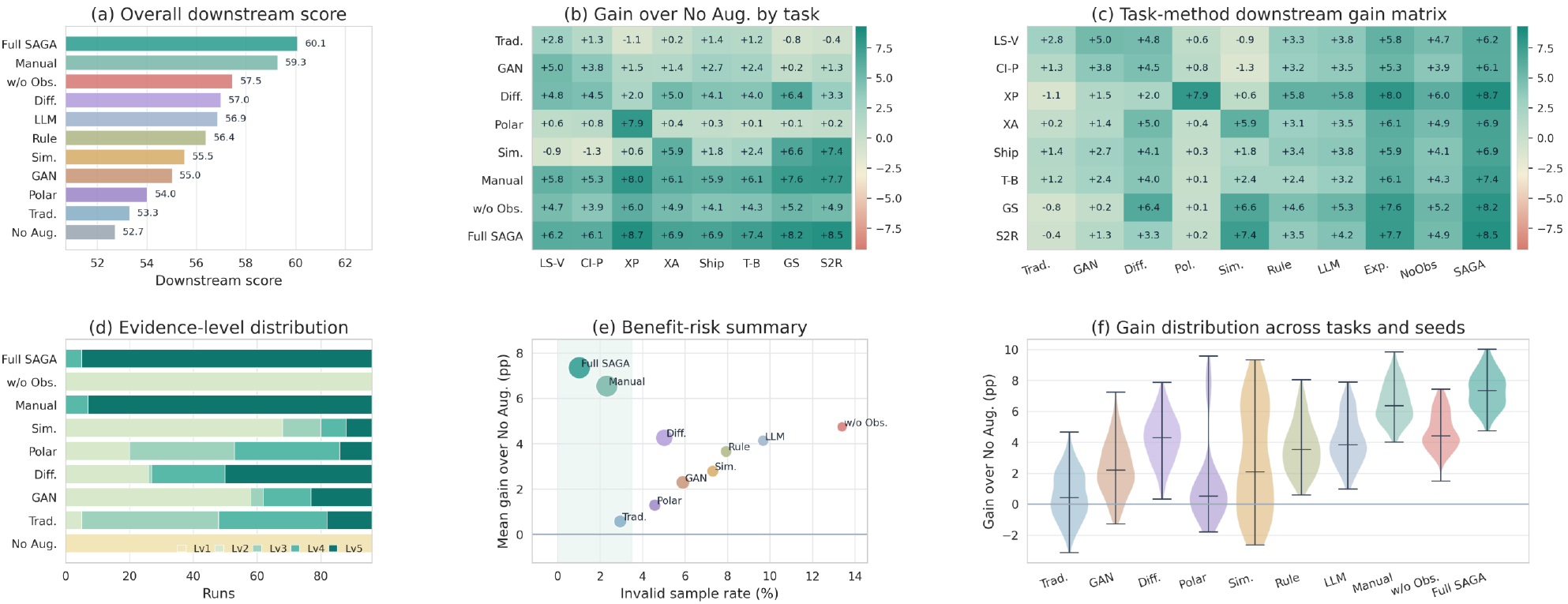}
    \caption{Downstream augmentation benefit across eight SAR task groups. Fixed augmentation strategies perform well only in their favorable scenarios, whereas Full SAGA adaptively selects augmentation skills and achieves the strongest average downstream score with the lowest invalid-sample rate among augmentation methods.}
    \label{fig:exp5_downstream}
\end{figure*}

\begin{table}[t]
\centering
\caption{Overall downstream augmentation results over 8 tasks $\times$ 12 seeds.}
\label{tab:exp5_overall}
{\scriptsize
\setlength{\tabcolsep}{2.0pt}
\renewcommand{\arraystretch}{1.08}
\begin{tabular*}{\columnwidth}{@{\extracolsep{\fill}}lccccc@{}}
\hline
Method & Score & Gain & Inv. & Cost & Lv5 \\
\hline
No Aug. & $52.7{\pm}5.1$ & $+0.0{\pm}0.0$ & 0.0 & 0.0 & 0.0 \\
Trad. & $53.3{\pm}6.0$ & $+0.6{\pm}1.6$ & 2.9 & 1.0 & 14.6 \\
GAN & $55.0{\pm}6.3$ & $+2.3{\pm}1.7$ & 5.9 & 3.1 & 19.8 \\
Diff. & $57.0{\pm}5.3$ & $+4.3{\pm}1.5$ & 5.0 & 6.8 & 47.9 \\
Polar & $54.0{\pm}6.1$ & $+1.3{\pm}2.7$ & 4.6 & 4.4 & 10.4 \\
Sim. & $55.5{\pm}3.7$ & $+2.8{\pm}3.4$ & 7.3 & 7.7 & 8.3 \\
Rule & $56.4{\pm}5.2$ & $+3.7{\pm}1.6$ & 7.9 & 4.7 & 7.3 \\
LLM & $56.9{\pm}5.1$ & $+4.1{\pm}1.6$ & 9.7 & 5.3 & 5.2 \\
Expert & $59.3{\pm}4.7$ & $+6.6{\pm}1.3$ & 2.3 & 5.8 & 92.7 \\
w/o Obs. & $57.5{\pm}5.2$ & $+4.7{\pm}1.2$ & 13.4 & 6.4 & 0.0 \\
Full SAGA & $\mathbf{60.1{\pm}4.5}$ & $\mathbf{+7.4{\pm}1.2}$ & \textbf{1.0} & 7.1 & \textbf{94.8} \\
\hline
\end{tabular*}}
\vspace{0.5mm}
\begin{flushleft}
\footnotesize
Inv. denotes invalid-sample rate. Lv5 denotes evidence-qualified downstream improvement rate.
\end{flushleft}
\end{table}

\begin{table}[t]
\centering
\caption{Paired significance tests over matched task-seed pairs. Confidence intervals and $p$-values are computed using a task-aware paired bootstrap. Task-wise Best Fixed denotes the strongest fixed-skill augmentation baseline selected separately for each task group.}
\label{tab:exp5_significance}
{\scriptsize
\setlength{\tabcolsep}{2.2pt}
\renewcommand{\arraystretch}{1.08}
\begin{tabular*}{\columnwidth}{@{\extracolsep{\fill}}lccc@{}}
\hline
Comparison & Diff. & 95\% CI & $p$ \\
\hline
SAGA vs Expert & +0.81 & [0.60, 1.01] & $<10^{-4}$ \\
SAGA vs Task-wise Best Fixed & +1.68 & [1.40, 1.95] & $<10^{-4}$ \\
SAGA vs w/o Observer & +2.62 & [2.34, 2.90] & $<10^{-4}$ \\
\hline
\end{tabular*}}
\end{table}

\subsubsection{Results}

Table~\ref{tab:exp5_overall} and Fig.~\ref{fig:exp5_downstream} show that Full SAGA obtains the strongest average downstream score, improving over No Augmentation by 7.4 percentage points on average. Fixed methods provide benefits only in favorable settings: Fixed GAN is effective for low-shot vehicle ATR, Fixed Polar is effective for cross-polarization ATR, Fixed Simulation is effective for cross-angle and sim-to-real tasks, and Fixed Diffusion is competitive for full-polarization imbalance, ship detection, and target-background scenes.

Full SAGA achieves performance comparable to, and slightly higher than, the manual expert policy under the same benchmark protocol. This result should not be interpreted as evidence that SAGA universally outperforms human experts. Rather, it indicates that an evidence-gated adaptive policy can match a fixed expert policy on this benchmark while maintaining a lower invalid-sample rate. The SAGA w/o Observer variant still improves the downstream score, but its Lv5 rate is 0.0 because Lv5 requires observer and leakage gates. This supports the distinction between raw downstream improvement and evidence-qualified augmentation benefit.

\begin{figure}[!t]
    \centering
    \includegraphics[width=\linewidth]{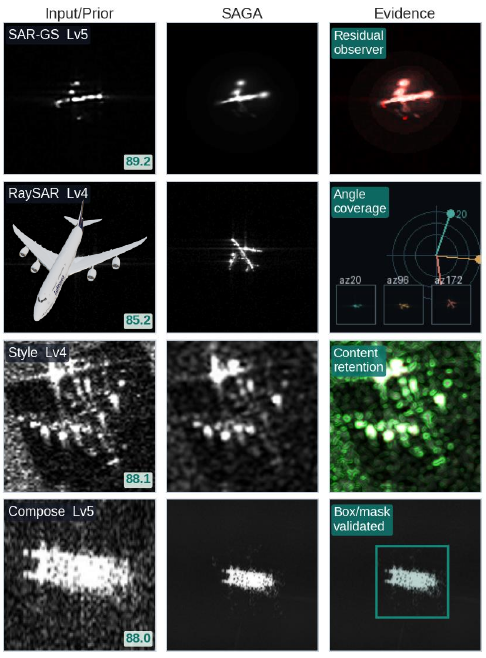}
    \caption{Qualitative SAGA cases. Each row shows the input or prior, the SAGA-generated result, and the corresponding evidence visualization. These examples illustrate that SAGA forms different decision chains under different task and data conditions. Since no downstream evaluator is executed inside Exp.6, the evidence level should be capped at Lv4 unless linked to Exp.5 downstream evaluation.}
    \label{fig:exp6_cases}
\end{figure}

The improvement of Full SAGA comes with a higher normalized augmentation cost than several fixed baselines. This is expected because SAGA may select heavier diffusion or simulation skills when the expected task benefit justifies the resource cost.

\subsection{Qualitative Case Analysis}

\subsubsection{Setup}

Exp.6 provides qualitative examples illustrating how SAGA exposes not only final augmented samples, but also selected skills, rejected alternatives, observer evidence, and provenance. The cases cover polarization-aware vehicle augmentation, sparse-view aircraft completion, ray-tracing aircraft simulation, cross-platform style migration, ship target-background composition, and observer-triggered repair.

\subsubsection{Results}

Fig.~\ref{fig:exp6_cases} illustrates several representative decision chains. For sparse-view aircraft, SAGA selects a view-completion strategy and reports coverage-based evidence. For ray-tracing aircraft simulation, it selects a simulation-oriented recipe and reports angle coverage. For cross-platform style migration, it emphasizes content retention rather than downstream performance. For ship target-background composition, it validates mask and box consistency.

These qualitative cases support traceability rather than serving as primary quantitative evidence. Therefore, we do not use subjective visual scores as main-text claims unless a blinded rater protocol and inter-rater agreement are provided. Since Exp.6 does not run downstream evaluators, its case-level evidence is capped at Lv4.

\subsection{Ablation Study}

\subsubsection{Setup}

Exp.7 evaluates the contribution of key SAGA modules by removing one component at a time: LLM intent parsing, schema validation, skill guardrails, benefit ranking, recipe DAG, observers, bounded repair, and policy memory. We report both component metrics and a composite diagnostic score. The composite score is used only as a compact visualization of degradation patterns and is not an independent benchmark metric.

The diagnostic composite score is computed as a fixed weighted average of normalized component metrics:
\begin{equation}
S_{\mathrm{comp}}=100\sum_k w_k\tilde{m}_k,
\end{equation}
where $\tilde{m}_k$ includes normalized intent, schema, skill, recipe, observer, invalid-selection, invalid-execution, invalid-sample, Lv5, and gain terms. Specifically, the invalid-rate terms are converted as $V=1-\mathrm{clip}(I/0.35,0,1)$ and the gain term is normalized as $G_{\mathrm{norm}}=\mathrm{clip}(G/8.6,0,1)$. Lv5 and Gain in this ablation benchmark are diagnostic metrics and are not directly comparable to the Exp.5 downstream benchmark summary.

\begin{table*}[t]
\centering
\caption{Ablation study. Composite is a diagnostic summary; interpretation should primarily rely on the individual failure signatures. Lv5 and Gain are measured on the ablation benchmark and are not directly comparable to the Exp.5 downstream benchmark summary.}
\label{tab:exp7_ablation}
{\scriptsize
\setlength{\tabcolsep}{2.3pt}
\renewcommand{\arraystretch}{1.06}
\begin{tabular*}{\textwidth}{@{\extracolsep{\fill}}lccccccccccc@{}}
\hline
Variant & Intent & Schema & Skill & Recipe & Obs. & Inv. Sel. & Inv. Exec. & Inv. Samp. & Lv5 & Gain & Comp. \\
\hline
Full SAGA & \textbf{95.0} & \textbf{96.0} & \textbf{94.0} & \textbf{96.0} & \textbf{91.0} & \textbf{2.0} & \textbf{3.0} & \textbf{5.0} & \textbf{69.0} & \textbf{7.25} & \textbf{90.4} \\
w/o Intent & 74.0 & 94.0 & 80.0 & 88.0 & 81.0 & 10.0 & 9.0 & 11.0 & 43.0 & 6.10 & 75.9 \\
w/o Schema & 93.0 & 58.0 & 82.0 & 79.0 & 75.0 & 12.0 & 16.0 & 15.0 & 37.0 & 5.88 & 68.0 \\
w/o Guard & 94.0 & 94.0 & 83.0 & 73.0 & 68.0 & 18.0 & 21.0 & 23.0 & 30.0 & 5.55 & 64.1 \\
w/o Rank & 94.0 & 95.0 & 86.0 & 91.0 & 84.0 & 7.0 & 7.0 & 9.0 & 47.0 & 5.58 & 80.7 \\
w/o Recipe & 94.0 & 95.0 & 88.0 & 69.0 & 77.0 & 7.0 & 24.0 & 16.0 & 36.0 & 5.97 & 70.5 \\
w/o Observer & 95.0 & 96.0 & 92.0 & 94.0 & 45.0 & 4.0 & 6.0 & 25.0 & 22.0 & 5.67 & 73.1 \\
w/o Repair & 95.0 & 96.0 & 93.0 & 95.0 & 78.0 & 3.0 & 5.0 & 13.0 & 52.0 & 6.55 & 83.7 \\
w/o Memory & 95.0 & 96.0 & 89.0 & 93.0 & 87.0 & 5.0 & 5.0 & 8.0 & 58.0 & 6.33 & 84.9 \\
\hline
\end{tabular*}}
\end{table*}

\subsubsection{Results}

Table~\ref{tab:exp7_ablation} shows that SAGA reliability arises from the interaction of multiple modules. Removing guardrails produces the largest composite drop and greatly increases invalid selection and invalid execution, indicating that skill compatibility checking is central to safe planning. Removing schema validation sharply reduces schema accuracy, confirming that dataset interpretations cannot be trusted without validator-gated grounding. Removing the recipe DAG primarily harms execution reliability, while removing observers sharply increases invalid sample rate. Removing benefit ranking has a smaller reliability drop but reduces downstream gain, suggesting that utility optimization and safety controls affect different parts of the system.

Repair and memory have smaller but consistent effects. Bounded repair mainly reduces residual invalid samples, while memory improves cross-case stability and downstream gain. These results support the view that SAGA is not a loose collection of tools; its performance depends on schema grounding, guardrails, recipe execution, observer evidence, repair, and memory working together.

\subsection{Summary}

Across the seven experiments, SAGA improves agentic reliability and downstream augmentation utility under controlled SAR augmentation benchmarks. Exp.1 shows that SAGA converts heterogeneous dataset formats into validator-gated schemas. Exp.2 shows that SAGA improves skill planning and avoids infeasible skill invocation compared with keyword, rule-only, LLM-only, and ReAct-style baselines. Exp.3 shows that recipe-centric execution provides auditability and reproducibility beyond ad hoc tool invocation. Exp.4 shows that observer-driven validation and bounded repair reduce invalid exports without claiming open-ended self-correction. Exp.5 shows that SAGA-selected augmentation achieves stronger average downstream utility than fixed augmentation strategies and comparable or slightly stronger performance than a manual expert policy under the same benchmark protocol. Exp.6 demonstrates traceable case-level behavior, and Exp.7 confirms that the main SAGA modules contribute complementary reliability and utility benefits.

\section{Conclusion}
This paper presented SAGA, a schema-grounded and benefit-aware agent framework for SAR data augmentation. SAGA addresses heterogeneous SAR datasets and diverse augmentation objectives by integrating deterministic dataset profiling, validator-gated schema compilation, LLM-assisted but rule-guarded planning, recipe-centric execution, observer-driven evaluation, bounded repair, and evidence-based policy memory. By separating flexible semantic proposal from deterministic validation and execution, SAGA provides a reproducible and evidence-aware workflow for selecting, executing, and assessing SAR augmentation strategies. Future work will expand the skill library, strengthen downstream evaluation across larger SAR benchmarks, and investigate memory-based planning under more diverse data distributions.

\section*{CRediT Authorship Contribution Statement}
Xuanting Wu: Conceptualization, Methodology, Software, Validation, Writing - original draft, Investigation, Formal analysis, Visualization. Fan Zhang: Supervision, Resources, Project administration, Funding acquisition. Fei Ma: Conceptualization, Investigation, Resources, Writing - review \& editing, Supervision, Project administration, Funding acquisition. Ling Guan: Data curation, Resources, Supervision, Project administration. Guochun Ma: Methodology, Software, Investigation. Yongsheng Zhou: Data curation, Supervision, Funding acquisition.

\section*{Declaration of Competing Interest}
The authors declare that they have no known competing financial interests or personal relationships that could have appeared to influence the work reported in this paper.

\section*{Data Availability}
The data that support the findings of this study are from private datasets and cannot be made publicly available.

\section*{Acknowledgements}
This work was supported in part by the National Natural Science Foundation of China under Grant Nos. 62271034 and 62331026, the Natural Science Foundation of Shandong Province under Grant ZR2024ZD19, and the Fundamental Research Funds for the Central Universities under Grant ZY2610.

\vfill

\end{document}